# Is the charge determined via shot noise measurements unique?

M. Dolev[1], Y. Gross[1], Y. C. Chung[2], M. Heiblum[1], V. Umansky[1], and D. Mahalu[1]

[1] *Braun Center for Submicron Research, Dept. of Cond. Matter Physics, Weizmann Institute of Science, Rehovot 76100, Israel*

[2] *Department of Physics, Pusan National University, Busan 609-735, Korea*

**ABSRACT**

Charged excitations in the fractional quantum Hall effect are known to carry fractional charges, as theoretically predicted and experimentally verified. Here we report on the dependence of the tunneling quasiparticle charge, as determined via highly sensitive shot noise measurements, on the measurement conditions, in the odd denominators states $v=1/3$ and $v=7/3$ and in the even denominator state $v=5/2$. In particular, for very weak backscattering probability and sufficiently small excitation energies (temperature and applied voltage), tunneling charges across a constriction were found to be significantly higher than the theoretically predicted fundamental quasiparticle charges.

Odd denominator fractional quantum Hall effect (FQHE) states [1, 2], whose quasiparticles are expected to possess fractional statistics, have been already the focus of extensive studies [3]. However, more recently, particular attention was devoted to the even denominator fractional state $v=5/2$ [4], which is conjectured to be described by a Pfaffian wavefunction [5-7], mainly due to the expectation that its quasiparticles carry a charge $e/4$ and obey non-abelian statistics [5, 8-10]. As such, these quasiparticles may be useful for topological quantum computation [11-13]. An important step in the experimental study of the FQHE states is the determination of the quasiparticle charge. While the fundamental quasiparticle charge in the bulk for a fractional state is expected to be unique, the charge that tunnels between two counter propagating edges-channels might depend on the measurement conditions. Here we study the uniqueness of the tunneling charge, and search for conditions under which it is equal to the expected fundamental quasiparticle charge.

Most charge measurements detect charges that tunnel either across a narrow constriction, via shot noise measurements [14-18], conductance measurements [19] and interference [20, 21], or charges that tunnel into localized states in the bulk [22], which, in all cases, are not guaranteed to be equal to the fundamental quasiparticle charge in the bulk. Moreover, the excitation energy (applied voltage and temperature) is also expected to affect the tunneling charge. However, since charges that tunnel between edges can only be integer multiples of the fundamental charge, the smallest measured charge sets an upper bound for the fundamental charge. For example, a measurement in the $v=5/2$ state of a charge $e/4$ rules out $e/2$ fundamental charges in this state. We present here measurement results of low frequency shot noise generated by partitioning. This work was motivated by our attempt to improve the accuracy of our previous measurements and tighten the data points with more sensitive measurements, thus allowing us to determine the charge in a previously inaccessible low energy and very weak backscattering regimes – where shot noise is excessively small. Our new measurements revealed an interesting dependence of the conductance and the tunneling charge on the energy and the transmission probability. The universality of these results was tested in a few fractional states, $v=1/3$, $v=7/3$, and $v=5/2$, where in all cases the conductance behaved in a highly characteristic way and the tunneling charge was found to be unexpectedly large in the very low energy and weak backscattering limits.

Measurements in the 2$^{nd}$ Landau level ($v=5/2$, $v=7/3$) were conducted on four different samples, which were fabricated on two different GaAs-AlGaAs heterostructures. The first (second) heterostructure, on which samples #1, #2 and #3 (sample #4) were fabricated, embedded a 2DEG confined in a 30nm wide quantum well (described by Umansky *et al*. [23]) with an areal electron density $2.9\times10^{11}$cm$^{-2}$ ($3.2\times10^{11}$cm$^{-2}$) and a low temperature mobility of $29\times10^{6}$cm$^{2}$/V-s ($30.5\times10^{6}$cm$^{2}$/V-s) - both measured in the dark. Hall measurement, taken in the ungated area of sample #1, is shown in Fig. 1a. Five significant fractional states, $v=11/5, 7/3, 5/2, 8/3,$ & $14/5$, are prominent (with $R_{xx}\sim0$ for $v=5/2$). Current was restricted by mesa etching, with the narrowest part being 5μm wide (see Fig. 1b). A metallic (15nm PdAu / 15nm Au) split-gate was deposited in the narrow part of the mesa; with split-gate separations 1μm, 1.2μm and 1.4μm. Studies in the 1$^{st}$ Landau level ($v=1/3$) were carried on sample #5, fabricated on a heterostructure embedding a 2DEG with an areal electron density $1.1\times10^{11}$cm$^{-2}$ and low temperature (dark) mobility $2\times10^{6}$cm$^{2}$/V-s, with split-gate separation of 350nm. Backscattering of edge channels was achieved by applying negative voltage to the split-gate with respect to the 2DEG. Note that relatively small gate voltages were needed, which in general only partly depleted the 2DEG under the gate (e.g., the gate voltage range used for bulk filling factor $v$=5/2 in sample #1 was $V_g$=-0.08V….-1.115V, while the full depletion voltage was -1.55V); however, the measured shot noise was universally found to depend on the transmission probability and not on the actual shape of the partitioning barrier (see for example Refs. [16-18, 24]).

Determination of charge from shot noise measurements was done via a rather simple analysis that has proven to be successful in determining the tunnelling quasiparticle charges in a variety of filling factors [14, 16-18, 24, 25]. The correctness of the analysis was validated in these earlier works by verifying the theoretically expected fundamental quasiparticle charge at sufficiently high temperatures, when the behavior is 'single-particle-like' and the conductance is linear. Charges $e$/3, $e$/5, $e$/7, $e$/4 were measured at $v$=1/3, 2/5, 3/7, 5/2 respectively, and an electron charge was measured in the integer regime. As will be discussed below, more advanced approaches, which rely on modeling the edge channel as a chiral Luttinger liquid (CLL) [26, 27], were found not to be applicable for our configuration. Our analysis assumed stochastic partitioning of shot-noise-free current carried by independent charged particles, emanating from a reservoir at finite temperature $T$. The partitioning leads to a binomial distribution of the partitioned particles [14-18, 28-31]. For multiple channel transport, partitioning the $i^{th}$ channel, which flows along the boundary between filling factors $v_i$ and $v_{i-1}$, leads to finite temperature low frequency spectral density of the current shot noise, $S^i(V_{sd},\omega\sim0,T)$, with the partitioned quasiparticle charge $e^*$ [29, 30]:

$$S^i(V_{sd},0)_T = 2e^*V_{sd}\Delta g_i\, t_{v_i-v_{i-1}}(1-t_{v_i-v_{i-1}})\left[coth(\frac{e^*V_{sd}}{2k_BT})-\frac{2k_BT}{e^*V_{sd}}\right]+4k_BTg\,, \qquad (1)$$

where $V_{sd}$ is the applied DC excitation voltage, $\Delta g_i = g_i - g_{i-1}$, with $g_j = \nu_j e^2/h$, $t_{\nu_i - \nu_{i-1}} = \frac{g - g_{i-1}}{\Delta g_i}$ is the transmission probability of the $i^{th}$ channel, and $g$ the two-terminal (Hall) conductance. In order to determine $\Delta g_i$ and $t_{\nu_i - \nu_{i-1}}$, the *next-lower-lying-state i*-1, which traverses freely through the partitioning constriction, must be identified. Note that when channels 1 to $i$-1 traverse with unity probability, while the $i^{th}$ channel is fully reflected, the two-terminal conductance, which is quantized at $g_{i-1} = \nu_{i-1} e^2/h$, is mostly current independent and the traversing current is shot-noise-free (see [18] for more details).

Pinching the constriction to fully reflect the $i^{th}$ channel, being it $\nu$=5/2 or $\nu$=7/3, the next-lower-lying state was identified in al cases to be $\nu$=2. For $\nu$=1/3 the lower-lying state is vacuum. Focusing on the dependence of the transmission coefficient on the excitation voltage, a repeatable behavior was observed. At high transmissions of the constriction (at small negative voltage on the split-gate), the differential conductance had a 'mound-like' dependence on the excitation voltage (a maximum at zero excitation voltage), while at low transmissions the conductance exhibited a 'valley-like' dependence (a minimum at zero excitation voltage). Such behavior was universally observed in the past, for all filling factors (integer or fractional) [18, 24, 32], being a characteristic of an induced scattering potential by a finite size potential formed, as an example, by the split-gate (with complete or partial depletion under the gate). Such non-linear behavior, and in particular the 'mound-like' dependence of the conductance, is not expected by a theory that models the edge channel by a CLL with a point-like scatterer - since it disregards realistic effects such as the spatial size of the scatterer and the dependence of the scatterer potential height on the applied excitation voltage. Hence, available theoretical predictions for the shot noise are also not applicable for the analysis of our experiments [24, 26, 27]. We would like to stress that a charge *e* was always observed in the integer regime by employing Eq. 1. This was independent if the conductance was 'mound-like' or 'valley-like'.

Shot noise and transmission were measured as function of the excitation voltage $V_{sd}$ and split-gate voltage $V_g$, for bulk filling factors $v=5/2$, $v=7/3$ and $v=1/3$; with tunnelling charge deduced from Eq. 1. In the weak and strong backscattering regimes, in all three filling factors, both *g* and $S^i(V_{sd},0)$ exhibited two distinct regions in the excitation voltage at ~10mK. At low excitation voltage a 'non linear characteristic' with voltage dependent *g*, was obtained, accompanied with a large slope in $S^i(V_{sd},0)$, which corresponds to a large charge. At higher excitation voltage a rather 'linear characteristic' and a significantly lower charge were obtained (see for example Figs. 1c & 1d, measured at bulk filling factor $v=5/2$).

Focusing on the range of low excitation voltage an interesting dependence of the deduced tunnelling charge on the average transmission probability was found. At $v=5/2$, at high - 'mound-like' - transmission (*t*~0.9), the tunneling (backscattered) charge was substantially higher than the theoretically predicted fundamental quasiparticle charge $e^*=e/4$ (Fig. 1c & 2a). At intermediate values of the transmission probability (*t*~0.4-0.9), with the conductance almost independent of excitation voltage, the tunnelling charge was very close to $e^*=e/4$ (Fig. 2b). At lower - 'valley-like' - transmission, the tunnelling charge increased towards $e^*=e$ (Fig. 2c). While the latter result is expected, since the filling factor within the constriction approaches $v=2$, thus enabling only tunneling of electrons, the high transmission results were not expected. Figure 2d summaries the tunnelling charge evolution as function of the average transmission $\langle t_{5/2-2}\rangle$, as measured with four different samples, in the low excitation voltage regime. A similar dependence of the tunnelling charge was also obtained at bulk filling factor $v=7/3$, as measured with sample #1. Two examples are shown in Figs. 3a & 3b (the somewhat higher excitation voltage was due to the very small excitation current – only 1/7 of the total current - carried by the 7/3 channel). The evolution of the tunnelling charge as function of $\langle t_{7/3-2}\rangle$ shown in Fig. 3c, resembling that for $v=5/2$, with $e^* \sim e/3$ obtained only at intermediate transmissions - where the transmission is nearly independent of excitation voltage.

For the $v=1/3$ state, we present only two examples of noise measurements, performed in the high range of the transmission $t_{1/3}$ with 'mound-like' and flat dependence (sample #5). A bimodal tunnelling charge is observed in Fig. 4a for a 'mound-like' transmission, with $e^*=e$ in the small excitation voltage range and $e^* \sim e/3$ at higher excitation voltages. At a somewhat lower, but a rather voltage independent transmission, the tunnelling charge fits well $e^*=e/3$ over a wide range of excitation voltage (Fig. 4b). In the range of lower transmission ($t_{1/3}<0.3$, not shown here), which had been already explored by Griffiths *et al*. [25], the backscattered charge approached $e$ - as expected for quasiparticles traversing a rather opaque barrier.

The temperature dependence of the shot noise for high transmission at bulk filling factor $v=5/2$ is shown in Fig. 5. The mixing chamber was heated (via a resistor) in the range 10-85mK, with the electron temperature monitored via the thermal noise, $S(0)_T=4k_BTg$. With an open constriction, the longitudinal resistance increased weakly with temperature, however the backscattering through the bulk was less than 0.1% even at 85mK. This backscattered current was shot noise free, and thus was subtracted from the impinging excitation current. Since the shot noise was substantially lower at the elevated temperatures, the range of the excitation voltage was extended $V_{sd}$=0-50μV. As evident from Figs. 5a & 5b the differential transmission coefficient was fairly constant with the excitation voltage and the quasiparticle charge diminished. The tunnelling charge approached $e^*=e/4$ as the temperature reached ~75mK and saturated thereafter (Fig. 5c).

Our shot noise measurements demonstrate that the tunnelling charge through a narrow constriction - that serves to partition the incoming current - is not unique and depends on the specific details of the tunneling barrier and the energetics. In our experiments the low energy charge (yet with $eV>>k_BT$) presented a rather complex evolution moving between two extreme values. A smaller value - being close to the theoretically expected fundamental charge – is measured at intermediate backscattering probabilities (or at higher excitation voltage or temperature), in which the transmission is either constant or exhibits a weak "valley-like" behavior. A larger charge, approaching $e$, for extremely weak backscattering and measured in the limit of low excitation voltage and temperature. Here the transmission is found to exhibit a "mound-like" behavior – in contradiction to the CLL prediction. The enhancement of the charge can be attributed to backscattering of a mixture of integer multiples of the fundamental charge ('bunching'). This assumption is strengthened by the observation of the lower measured charge at higher excitation voltage or temperature, which can be attributed to the dissemination of the higher charge into individual fundamental charges. This hypothesis finds support from the expectation that in states belonging to the Jain's series, namely, $v=p/(2np+1)$, with $n$ and $p$ integers, and in the presence of multiple edge channels, weak backscattering of bunched quasiparticles dominates at low temperatures [33, 34] (although the fractions $v=5/2$, $v=7/3$ and $v=1/3$ were not addressed). A similar larger charge is also measured when the backscattering of the relevant edge channel is strong. However, bunching in this limit is theoretically expected [28]. Our findings are not only theoretically intriguing but have implications on a variety of proposed experiments, such as proposed interference experiments designed to test the statistical nature of quasiparticles [35-38], as they rely on weak backscattering of quasiparticles. In light of our findings, the importance of charge determination prior to such measurements is evident.

**Acknowledgments**

We thank Ady Stern, Yuval Gefen, Bernd Rosenow, Dario Ferraro, and Maura Sassetti for helpful discussions. We thank the partial support of the Israeli Science Foundation (ISF), the Minerva foundation, the German Israeli Foundation (GIF), the German Israeli Project Cooperation (DIP), the European Research Council under the European Community's Seventh Framework Program (FP7/2007-2013) / ERC Grant agreement # 227716, the US-Israel Bi-National Science Foundation. YC was supported by the Korea Science and Engineering Foundation (KOSEF) grant funded by the Korean government (MEST, No. 2009-006003).

**Figure captions**

**Figure 1:** Hall measurements, the experimental setup, and conductance and noise measurements at $v$=5/2 state. **(a)** Hall measurement, taken in sample #1 at $T$=12mK. **(b)** A schematic description of the samples and the associated circuitry. Excitation DC current is driven via the source S, provided by a DC voltage $V$ and a large resistor in series. The AC voltage v is used to measure the two-terminal differential conductance. Drain voltage fluctuations (at D) is filtered with a LC resonant circuit, tuned to 910 kHz, amplified by a cooled preamplifier (A) attached to the 4.2K stage. The split-gate, controlled by voltage $V_g$, was tuned for the desired transmission probability. **(c), (d)** Measurements of shot noise and the corresponding different transmission $t_{5/2-2}$ at $T$=12mK, for a large range of excitation voltage.

**Figure 2:** Measurements at $v$=5/2 state of the transmission probability and the shot noise in the small $V_{sd}$ range, measured at $T$~10mK. **(a) - (c)** Spectral density for a few transmissions and the predicted spectral density for different partitioned quasiparticle charges (using Eq. 1). **(d)** The evolution of quasiparticle charge as a function of the average transmission probability $\langle t_{5/2-2} \rangle$, measured on four different samples.

**Figure 3:** Transmission coefficients and spectral density at $v$=7/3 state. **(a), (b)** The measured spectral density at two different transmission coefficients, and the predicted spectral density for different quasiparticle charges (using Eq. 1). **(c)** Backscattered charge as a function of the average transmission $\langle t_{7/3-2} \rangle$ at $v$=7/3 and $T$=12mK.

**Figure 4:** Transmission coefficient and spectral density at $v=1/3$ state. **(a), (b)** One-side spectral density at $v=1/3$ and $T=10$mK. The expected spectral density for charge $e^*=e$ and $e^*=e/3$ is plotted (solid lines using Eq. 2) on top of the data. The 'mound-like' behavior of the transmission is associated with the enhanced partitioned charge in the low impinging current range and a smaller charge, close to $e^*=e/3$, in the high current range. A 'flat' transmission coefficient, as in **(b)** results in $e^*=e/3$ in the full range of the excitation voltage (or impinging current).

**Figure 5:** The dependence of the backscattered charge on temperature measured at weak backscattering in the $v=5/2$ state. Two examples of the measured data: **(a)** at $T=46$mK with $e^* \sim 0.5e$, and **(b)** at $T=76$mK with $e^* \sim e/4$. **(c)** Evolution of the backscattered charge as a function of temperature. At temperatures higher than ~40mK a significant reduction of the charge is observed.

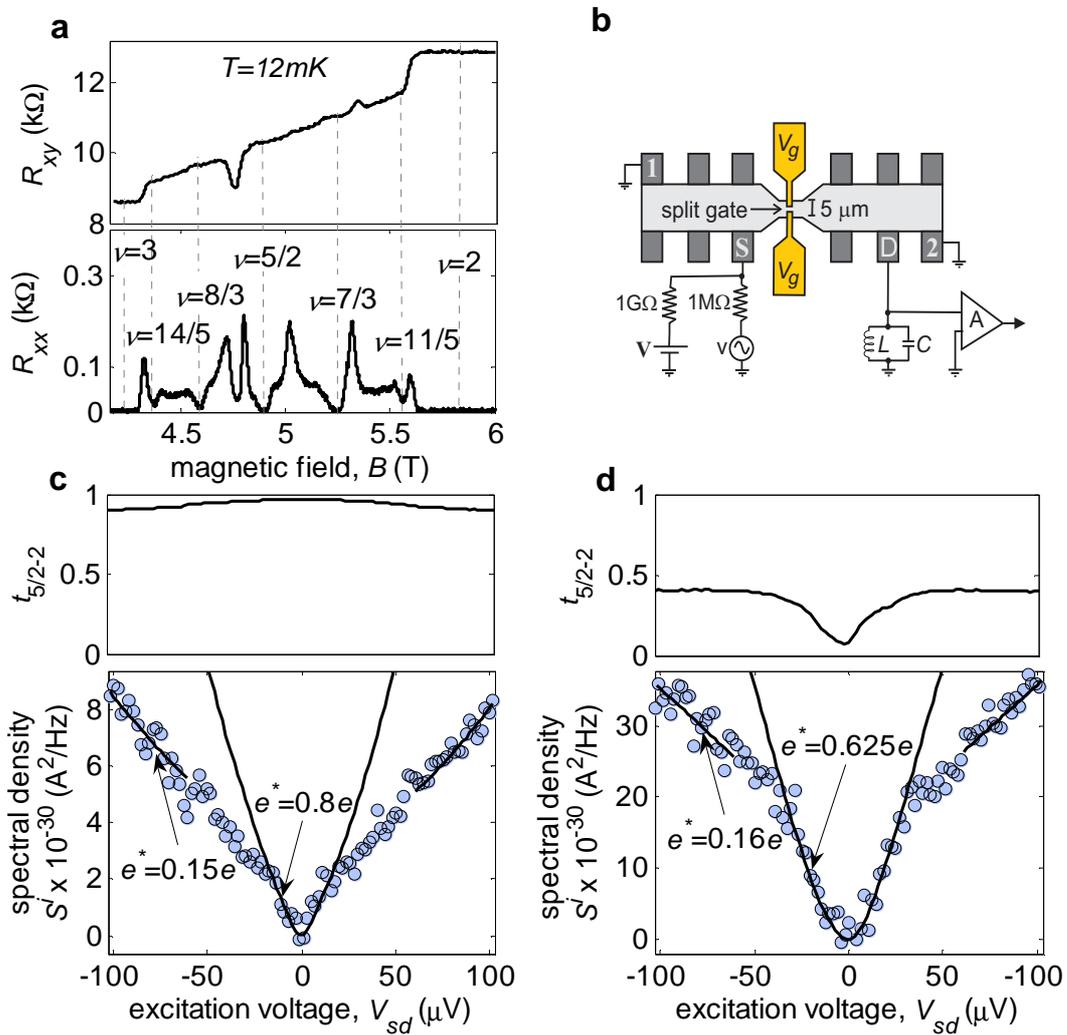

Figure 1

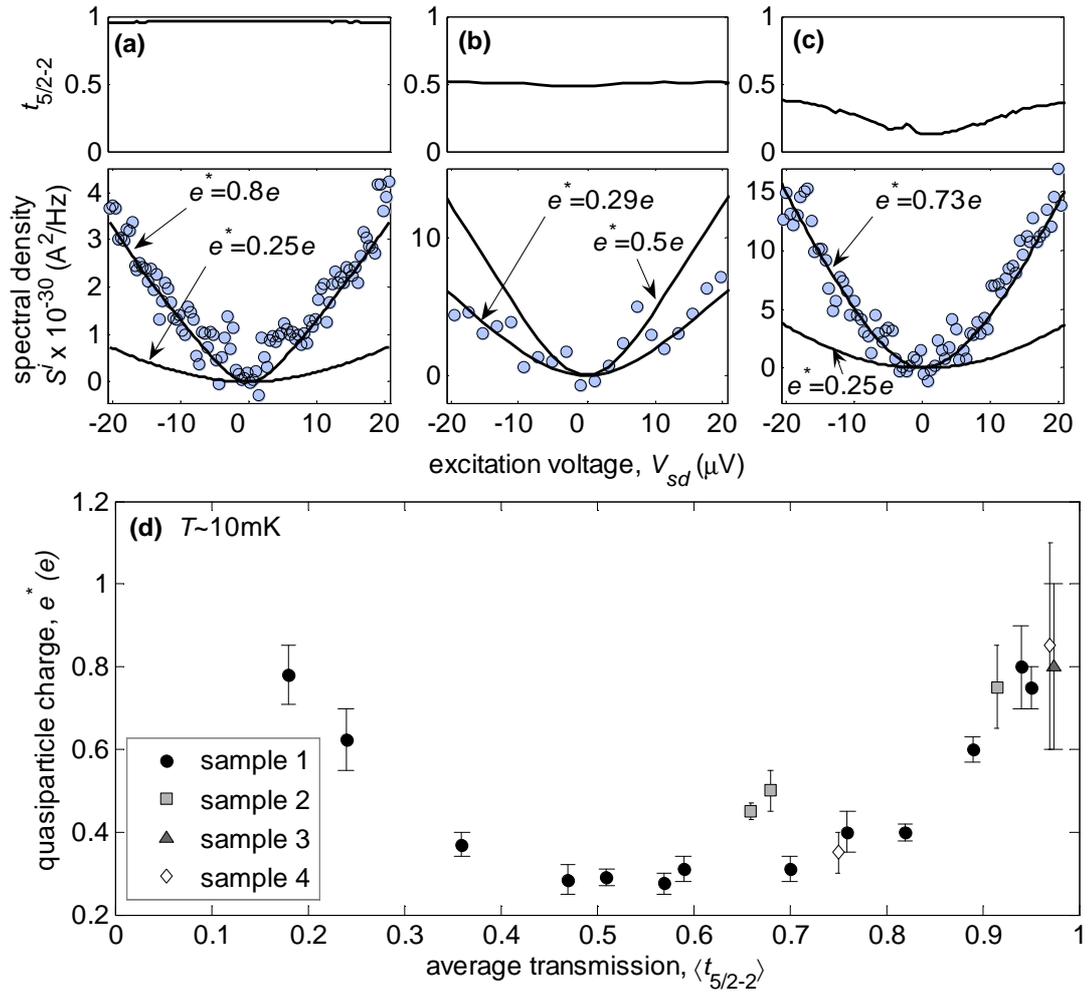

Figure 2

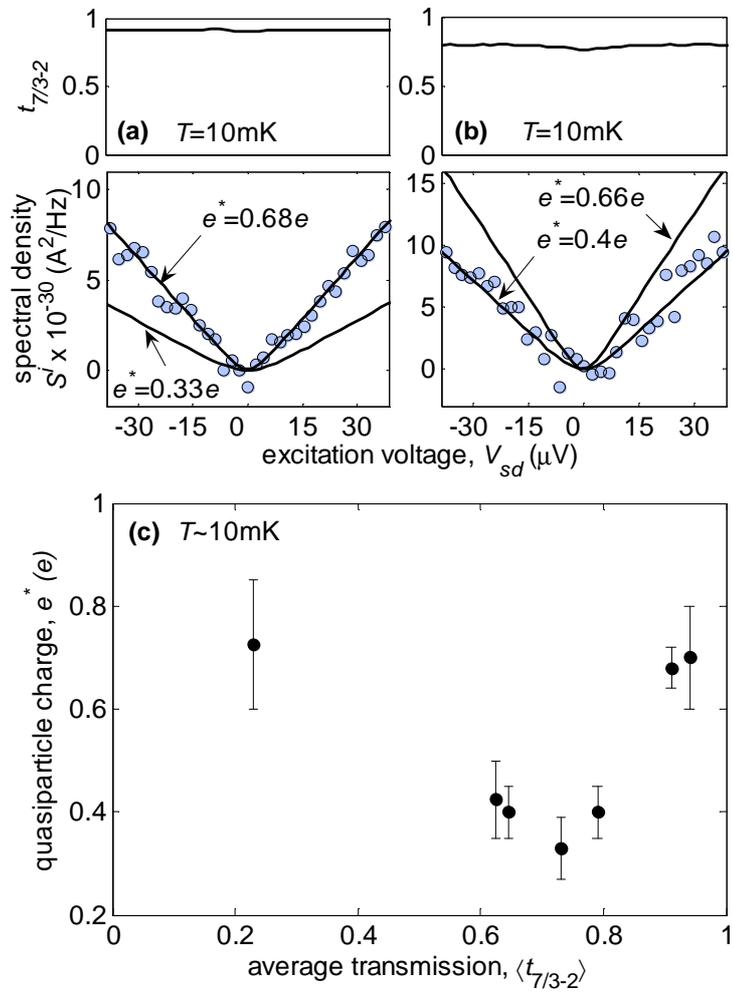

Figure 3

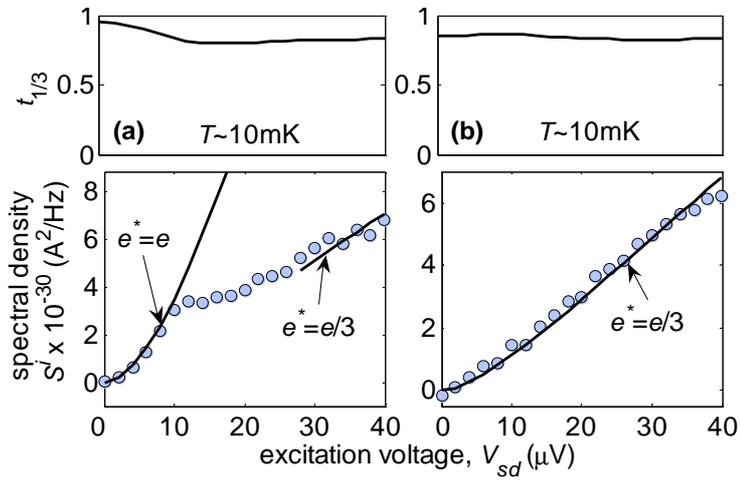

Figure 4

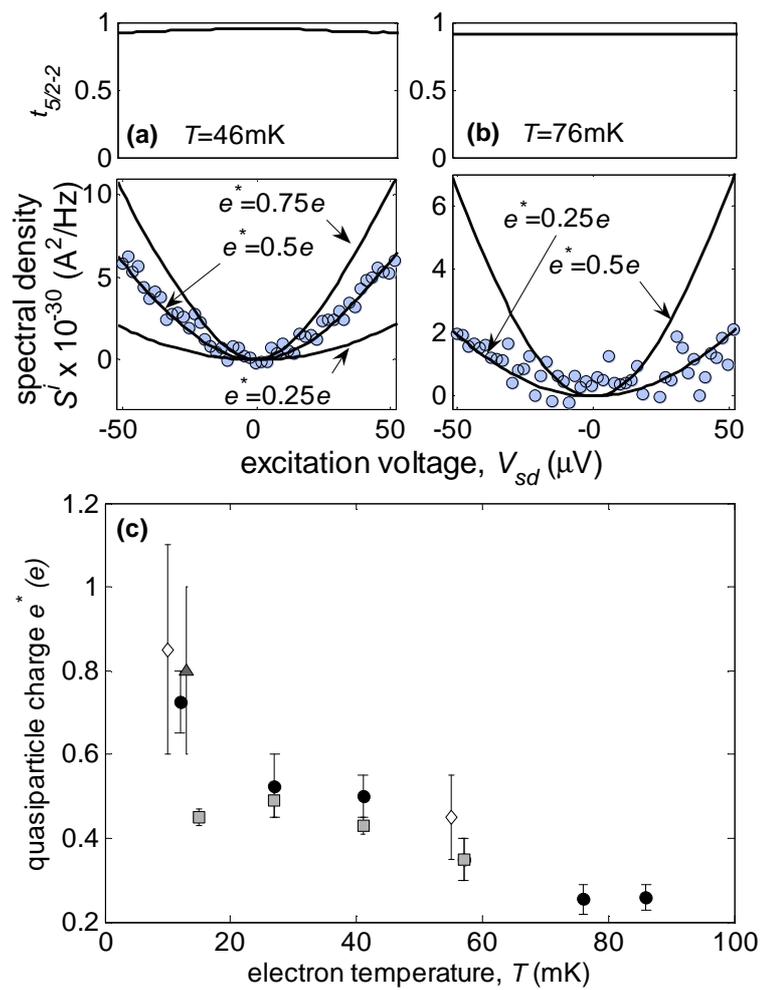

Figure 5